\documentclass[11pt,a4paper]{article}
\usepackage{epsfig, picinpar}
\usepackage{amssymb,amscd}
\newcommand{\rem}[1]{}

\title{A New Integrable Equation with Peakon  
Solutions}

\author{A. Degasperis\thanks{Dipartimento di Fisica,  
Universit\`{a} di Roma ``La Sapienza'', P.le A. Moro 2, 
00185 Roma, Italia. E-mail: 
antonio.degasperis@roma1.infn.it}, 
D.D. Holm\thanks{Theoretical Division and Center for 
Nonlinear Studies, Los Alamos National Laboratory, Los 
Alamos, NM 87545, USA. E-mail:dholm@lanl.gov} 
\& A.N.W. Hone\thanks{Institute of Mathematics \& Statistics, 
University of Kent, Canterbury CT2 7NF, UK. 
E-mail:anwh@ukc.ac.uk} 
}

\begin{document}
\renewcommand{\theequation}{\arabic{section}.\arabic{equation}}
\newcommand{\beq}{\begin{equation}}
\newcommand{\eeq}{\end{equation}}
\newcommand{\bea}{\begin{eqnarray}}
\newcommand{\eea}{\end{eqnarray}}
\maketitle

\begin{abstract}
We consider a new partial differential equation,
of a similar form to the Camassa-Holm shallow water wave equation, 
which was recently obtained by  
Degasperis and Procesi using the method 
of asymptotic integrability. 
We prove the exact integrability of the new
equation by constructing its Lax pair, and we explain its connection with a
negative flow in the Kaup-Kupershmidt hierarchy via a 
reciprocal transformation. The infinite sequence of conserved quantities    
is derived together with a proposed bi-Hamiltonian  
structure. 
The equation admits exact solutions in the form of a superposition of  
multi-peakons, and we describe the integrable 
finite-dimensional peakon 
dynamics and compare it with the analogous results for Camassa-Holm 
peakons.   
\end{abstract}

\section{Introduction}

Since its discovery \cite{ch} there has been a considerable 
amount of interest in the Camassa-Holm 
shallow-water equation, 
\beq 
u_t+2\kappa u_x+ \gamma u_{xxx}-u_{xxt}+3uu_x=2u_xu_{xx}+uu_{xxx}.
\label{eq:caholm} 
\eeq     
It was originally derived as an approximation to the incompressible Euler  
equations, and found to be completely integrable with a 
Lax pair and associated bi-Hamiltonian structure \cite{ch2}.   
It was subsequently shown to have a hodograph link to the KdV hierarchy 
and an interpretation within the framework of hereditary recursion operators
\cite{fuchss}.  In  \cite{DGH[2001]}, the CH equation with linear 
dispersion was shown to be one full order more accurate in asymptotic
approximation beyond Korteweg-de Vries (KdV) for shallow water waves. Yet,
it still preserves KdV's soliton properties such as complete integrability
via the inverse scattering transform (IST) method. 2+1-dimensional
analogues have also been studied 
\cite{hone3, kraenkel}. A particularly unusual feature of (\ref{eq:caholm}) 
is that in the dispersionless limit $\kappa\to 0$ it admits peaked 
solitons or peakons. The single peakon takes the form 
\beq 
u(x,t)=ce^{-|x-ct|}, \label{eq:1peakon} 
\eeq 
while the $N$-peakon solution is just a 
simple superposition, 
\beq 
u(x,t)=\sum_{j=1}^Np_j(t)e^{-|x-q_j(t)|}, 
\label{eq:Npeakon} 
\eeq 
where the canonical positions $q_j$ and momenta $p_j$ satisfy 
a completely integrable finite-dimensional Hamiltonian system.

The current work was motivated by the question of which equations of  
similar form to (\ref{eq:caholm}) are integrable. 
In the recent study \cite{dega} the method of asymptotic  
integrability was applied to 
the family of third order dispersive PDE conservation laws,
\beq 
u_t +c_0 u_x + \gamma u_{xxx} -\alpha^2 u_{xxt}
= (c_1 u^2 + c_2 u_x^2 + c_3 uu_{xx})_x,  
\label{eq:laws} 
\eeq
whose right-hand side is the derivative of a quadratic differential
polynomial. 
Within this family, only three equations that satisfy 
asymptotic integrability conditions up to third order 
are singled out, namely KdV 
($\alpha=c_2=c_3=0$), 
Camassa-Holm ($c_1=-\frac{3}{2}c_3/\alpha^2, c_2=c_3/2$) 
and one new equation ($c_1=-2c_3/\alpha^2, c_2=c_3$) which 
can be scaled to the following form:  
\begin{eqnarray}
u_t+u_x+6uu_x+u_{xxx}
-\alpha^2\left(u_{xxt}+\frac{9}{2}u_x u_{xx} 
+\frac{3}{2}uu_{xxx}\right)
=0
\,.
\label{eq:tdorig}  
\end{eqnarray}
(Unfortunately, the equation in \cite{dega} appears with
the typographical error $3/2\rightarrow2/3$.) 
However, by its nature the method of \cite{dega} only isolates 
necessary conditions for integrability and is not sufficient to prove it. 
In the following we demonstrate that 
the new equation (\ref{eq:tdorig}) is integrable 
by explicitly constructing its Lax pair. 

A simple observation that proved very useful in  our investigations 
was that  
by rescaling, shifting the dependent 
variable and applying a Galilean boost, the equation  
(\ref{eq:tdorig})  may be transformed to the dispersionless form 
\beq 
u_t-u_{xxt}+4uu_x=3u_xu_{xx}+uu_{xxx}. \label{eq:tdnodisp}
\eeq
but integrability is unaffected by such transformations.  
(We have set the parameter $\alpha=1$, but it will sometimes be 
convenient to reintroduce it when considering various limits.) 
The next key observation was that in this form the equation 
(\ref{eq:tdnodisp}) also has the single peakon (\ref{eq:1peakon}) 
as an exact travelling wave solution. This led us to consider the 
family of equations 
\beq
u_t-u_{xxt}+(b+1)uu_x=bu_xu_{xx}+uu_{xxx}. \label{eq:bfamily}
\eeq
for real parameter $b$, which includes both Camassa-Holm ($b=2$) and 
the new equation (\ref{eq:tdnodisp}) ($b=3$) as special cases. 
It turns out that all the equations in this family possess not just the 
peakon solution  (\ref{eq:1peakon}) but also the multi-peakon 
solutions (\ref{eq:Npeakon}). In the case of arbitrary $b$ the $p_j,q_j$ 
are not canonical variables, but satisfy the dynamical system 
\beq 
\dot{p}_j = 
-(b-1)\,\frac{\partial G_N}{\partial q_j},
\qquad  
\dot{q}_j  =  
\frac{\partial G_N}{\partial p_j}
,
\label{eq:ODEeqn}
\eeq  
where the generating function $G_N$ is 
\begin{equation} \label{G_N-def}
G_N = \frac{1}{2}\sum_{j,k=1}^{N} p_j p_k\, e^{-|q_j-q_k|} 
\,;
\end{equation}
this takes the canonical Hamiltonian form only in the special case $b=2$ 
(Camassa-Holm). 

The asymptotic analysis in \cite{dega} implies that $b=2,3$ should be 
the only possible integrable cases within the family (\ref{eq:bfamily}). 
In \cite{pick} a class of PDEs  
including all of the equations (\ref{eq:bfamily}) was tested for  
integrability using Painlev\'e analysis. However, 
because both Camassa-Holm equation (\ref{eq:caholm}) and the new equation 
(\ref{eq:tdnodisp}) are examples of integrable systems  
with algebraic branching in their solutions (the weak Painlev\'e property 
of \cite{weak}), they were explicitly excluded by the various 
Painlev\'e tests applied in \cite{pick}, and in fact all 
of the equations in 
that class failed the combination of tests.  
The authors of \cite{pick} noted that the (strong) Painlev\'e property 
is destroyed by changes of variables, and thus a transformation may be 
required before applying the test.  
For the case of Camassa-Holm a reciprocal 
transformation provides a link with a negative flow of KdV (see 
\cite{fuchss, hone1}), which restores the Painlev\'e property. 
In a forthcoming article  \cite{dhh} 
we shall generalize this transformation for the whole 
family (\ref{eq:bfamily}). Applying Painlev\'e analysis 
to the transformed family of equations then isolates the cases 
$b=2,3$ as the only candidates for integrability. 
       
In this note we restrict our attention to the new integrable  
equation (\ref{eq:tdnodisp}). We first present the associated linear system 
with a third order 
Lax operator. We explain the connection with a negative flow of the 
Kaup-Kupershmidt hierarchy \cite{pick2},  
and derive infinitely many conservation laws 
and a proposed bi-Hamiltonian structure. The dynamics of multi-peakon 
solutions, described by the finite-dimensional system (\ref{eq:ODEeqn}) in 
the particular case $b=3$, is also analysed.

\section{Reciprocal transformation}

\setcounter{equation}{0}
We consider the equation (\ref{eq:tdnodisp}) which we write in the form  
\beq
m_t+m_x u +3m u_x=0, \qquad m=\mathcal{L}u:=(1-\partial_x^2)u.
\label{eq:peak}
\eeq
In this form it is clear that 
we have the conservation law 
\beq
(m^{1/3})_t=-(m^{1/3}u)_x. \label{eq:con}
\eeq
The construction of the other conservation laws (infinitely many) will 
be presented 
in Section 4. 
 
Our strategy is to make use of a reciprocal transformation which maps  
between systems of conservation
laws \cite{rogers1, rogers3, rogers2},
since it is known that such transformations
can be used to connect systems with algebraic branching in their
solutions to equations which have only pole type singularities.
The relevant example here is the Camassa-Holm equation, which has
a reciprocal transformation to the first negative flow in the KdV hierarchy,
namely the equation
\beq
R\cdot V_T=0,
\label{eq:inv}
\eeq
where $R=\partial_X^2+4V+2V_X\partial_X^{-1}$ is the KdV recursion operator
(for more details see \cite{hone1} and references). Camassa-Holm
has weak Painlev\'{e} expansions with cube root branching \cite{pick},
while KdV
solutions have only double pole singularities. 

To define a reciprocal transformation
for the equation (\ref{eq:peak}) 
we introduce the dependent
variable
\beq 
p=-m^{1/3} 
\eeq
which is related to $u$ by 
\beq
p^3=(\partial_x^2-1)u
\,. \label{eq:rel}
\eeq
We then use the conservation law (\ref{eq:con}) in the form
\beq
p_t=-(pu)_x
\label{eq:con3}
\eeq
to define the new independent variables
$X,T$ via
\beq
dX=p\, dx-pu\,dt, \qquad dT=dt.
\label{eq:recip}
\eeq
By transforming the derivatives we find the new conservation law
\beq
(p^{-1})_T=u_X. \label{eq:ncon}
\eeq
Replacing $\partial_x$ by $p\partial_X$ in (\ref{eq:rel})
and using (\ref{eq:ncon}) to eliminate derivatives of $u$ leads to the
identity $u=-p(\log p)_{XT}-p^3$. Hence, (\ref{eq:ncon}) can be written in
terms of
$p$ alone, viz:   
\beq
(p^{-1})_T+(p(\log p)_{XT}+p^3)_X=0. \label{eq:receqn}
\eeq
It then turns out that this equation for $p$
can be written in conservation form in another way, namely
\beq
\left(-\frac{p_{XX}}{p} +\frac{p_X^2}{2p^2}-\frac{1}{2p^2}\right)_T
=\frac{3}{2}(p^{2})_X. 
\label{eq:ncon2}
\eeq

The equation (\ref{eq:ncon2}) is most conveniently written as the
system
\beq
\begin{array}{lcl}
V_T \, = \, \frac{3}{4}(p^{2})_X, & &  \\
pp_{XX}-\frac{1}{2}p_X^2+2Vp^2+\frac{1}{2} & = & 0, \end{array}
\label{eq:tdp}
\eeq
where the second equation above (Ermakov-Pinney) defines $V$. A direct
consequence of Ermakov-Pinney is
\beq
B\cdot p=0, \qquad B\equiv R\cdot\partial_X =\partial_X^3+4V\partial_X+2V_X.
\label{eq:sechamop}
\eeq
For a summary of results and references on the Ermakov-Pinney equation see
\cite{hone2}. Having obtained the equation (\ref{eq:receqn}) 
from (\ref{eq:peak}) by using the reciprocal transformation
(\ref{eq:recip}), one may check that the standard WTC
Painlev\'{e} test  \cite{wtc} is satisfied. In fact the original equation
(\ref{eq:peak})  admits expansions with square root branching, but such
generalized  (weak \cite{weak}) Painlev\'{e} expansions were specifically
excluded  from the classification of \cite{pick}. Further details of the 
Painlev\'{e} analysis will be given elsewhere \cite{dhh}.

\section{The Lax pair}

\setcounter{equation}{0}

Equation (\ref{eq:receqn}), which is equivalent to 
the system (\ref{eq:tdp}),  
is integrable since it admits the Lax pair
\beq
\begin{array}{rcc}
\psi_{XXX}+4V\psi_X+(2V_X-\lambda)\psi & = & 0, \\
\psi_T+\lambda^{-1}(p^2\psi_{XX}-pp_X\psi_X
-(pp_{XX}-p_X^2+\frac{2}{3})\psi) & = & 0.
\end{array}
\label{eq:tdplax}
\eeq
The third order operator above is the Lax operator of the Kaup-Kupershmidt
(KK) hierarchy \cite{pick2}, 
and so the system (\ref{eq:tdp}) may be considered as the
first negative flow in this hierarchy. 
As far as we are aware this system has not been studied before,
although it turns out to be an integrable generalization of
the Tzitzeica equation. 
(The constant $2/3$ in the $T$ part of 
(\ref{eq:tdplax}) can be gauged away, but allows for easy 
comparison with reference \cite{pick2}).

However, it is not the most general negative KK flow
derived from such a Lax pair with an inverse power of $\lambda$
in the $T$ part. Indeed, starting from a more general time evolution
of the form
\beq
\psi_T+\lambda^{-1}\left(W\psi_{XX}-\frac{1}{2}W_X\psi_X
+\frac{1}{6}(W_{XX}+16VW)\psi\right)=0,
\label{eq:wlax}
\eeq
which is a reduction of the time part of a 2+1-dimensional
non-isospectral KK hierarchy introduced in \cite{pick2},
the compatibility conditions of the third order Lax equation
with (\ref{eq:wlax}) yields the two conditions
\beq
V_T=\frac{3}{4}W_X, \qquad \tilde{B}\cdot W=0, \label{eq:compat}
\eeq
where
$ 
\tilde{B}=\partial_X^5+6(\partial_X V\partial_X^2+\partial_X^2V\partial_X)
+4(\partial_X^3V+V\partial_X^3)+32(\partial_XV^2+V^2\partial_X).
$ 
The recursion operator of the KK hierarchy is
\beq 
R_{KK}=B\partial_X^{-1}\tilde{B}\partial_X^{-1},
\label{eq:kkrec} 
\eeq 
with $B$ given by (\ref{eq:sechamop}); $B$ is
a Hamiltonian operator for the KdV, KK and Sawada-Kotera hierarchies.  
(see \cite{pick2} and references).

Thus we see that the more general inverse KK flow (\ref{eq:compat})
is a system of higher order than (\ref{eq:tdp}) (while the most general
first negative flow is $R_{KK}\cdot V_T=0$, of even higher order).
However, the
striking thing is that system (\ref{eq:tdp}) (or equivalently 
(\ref{eq:receqn}))
is a consistent reduction of (\ref{eq:compat}). This is a consequence of
the following remarkable operator identity:
\beq
\tilde{B}\cdot p^2=2(p\partial_X^{2}+5p_X\partial_X+10p_{XX}+16Vp)\cdot
B\cdot p.
\label{eq:factor}
\eeq
Thus for an arbitrary function $h(T)$ we may substitute
\beq 
W=p^2, \qquad V=-\frac{p_{XX}}{2p}+\frac{p_X^2}{4p^2}-\frac{h(T)}{4p^2}
\eeq 
into the $T$ part (\ref{eq:wlax})
of the Lax pair, and then upon scaling $h(T)=1$
(always possible for $h\not\equiv 0$) 
the compatibility condition of the Lax pair becomes (\ref{eq:tdp}).
This is consistent with the second condition of (\ref{eq:compat})
due to the identity (\ref{eq:factor}) together with $B\cdot p=0$ 
(\ref{eq:sechamop}).
In the special case  
$h(T)=0$, which is not
relevant to the reciprocal transformation of the equation
(\ref{eq:peak}), we can integrate the resulting system
to obtain
\beq
p(\log p)_{XT}+p^3+\tilde{h}(T)=0
\label{eq:tzit}
\eeq
with $\tilde{h}$ arbitrary. For $\tilde{h}\not\equiv 0$
we rescale and redefine $T$
so that $\tilde{h}=1$, in which case
(\ref{eq:tzit})
it is known as the Tzitzeica equation \cite{tzit},
or with dependent variable $\phi=\log p$ 
the Bullough-Dodd equation \cite{bull} 
\beq 
\phi_{XT}+e^{2\phi}+e^{-\phi}=0.
\eeq 
If $\tilde{h}=0$ we have Liouville's equation.

Finally we present the Lax pair for the original equation
(\ref{eq:peak}) (or equivalently (\ref{eq:tdnodisp})).
Setting
\beq 
\Psi=p^{-1}\psi, 
\eeq 
recalling that
$m=-p^3$,  
and transforming all derivatives in (\ref{eq:tdplax}),
we find that $\Psi$ satisfies 
\beq 
\begin{array}{rcc}  
\Psi_x-\Psi_{xxx}-m\lambda\Psi
& = & 0, \\ 
\Psi_t
+
\frac{1}{\lambda}\Psi_{xx}+u\Psi_x- 
\left(u_x+\frac{2}{3\lambda}\right)\Psi
& = &  
0.  
\end{array} 
\label{eq:nodisplax} 
\eeq 
(As before the final constant $2/(3\lambda)$ can be removed by a 
gauge transformation.) 

It is worth considering the effect of a Galilean boost and rescaling 
on the equation (\ref{eq:tdnodisp}), which restores some of the free 
parameters in (\ref{eq:laws}), yielding  
\beq 
m_t+c_0 u_x + \gamma u_{xxx}+ m_x u + 3mu_x=0, \quad m=u-\alpha^2 u_{xx}. 
\label{eq:dispfam}   
\eeq      
In that case, with a rescaled spectral parameter $\mu=\alpha^2\lambda$, 
the Lax pair is 
\beq 
\begin{array}{l}  
(1-\alpha^2\partial_x^2)\Psi_x
=
\mu\Big(m+\frac{c\,_0}{3}
+
\frac{\gamma}{3\alpha^2}
\Big)\Psi, \\  
\Psi_t + \frac{1}{\mu}\Psi_{xx}
+
\Big(u-\frac{\gamma}{\alpha^2}\Big)\Psi_x-u_x\Psi
=0. 
\end{array} 
\eeq 
In the dispersionless case, if $c_0=0=\gamma$ then up to 
scaling the equation 
(\ref{eq:dispfam}) is equivalent to the Riemann shock wave equation 
\beq 
u_t+uu_x=0 
\eeq 
in both the limits $\alpha^2\to 0$ and $\alpha^2\to\infty$, i.e. 
in the limit of both large and small wavenumbers.%
\footnote{Actually, the limit $\alpha^2\to\infty$ of (\ref{eq:dispfam}) in
the dispersionless case is
$\partial_x^2(u_t+uu_x)=0$.}

The generalized Tzitzeica equation (\ref{eq:receqn}) has a variety of 
interesting solutions, including solitons on a constant background and a 
reduction to a Painlev\'{e} III transcendent. We will treat these solutions  
in more detail in \cite{dhh}. Here we mention that the 
reciprocal transformation takes the one-soliton solution on constant
background to a soliton solution of the equation (\ref{eq:dispfam})
with dispersion  and vanishing at infinity, while the peakon
solution         (\ref{eq:1peakon}) can be obtained from the soliton in the
limit of  zero dispersion (although the transformation (\ref{eq:recip})
breaks  down for the peakon, when $m=-p^3$ is a Dirac delta function). 
   
\section{Conservation laws} 

\setcounter{equation}{0}

In deriving infinitely many conservation laws for 
the equation (\ref{eq:tdnodisp}), 
we first introduce the quantity 
\beq 
\rho=(\log\psi)_x
\,. \label{eq:rho}  
\eeq 
Then from the spatial part of the Lax pair (\ref{eq:nodisplax}) 
we have 
\beq 
\mathcal{L}\rho:=(1-\partial_x^2)\rho=3\rho\rho_x+\rho^3+\lambda m, 
\label{eq:spat} 
\eeq 
while (removing the $2/(3\lambda)$ term) 
the time part yields a conservation law for $\rho$, 
namely 
\beq 
\rho_t=j_x, \qquad j=u_x-u\rho-\lambda^{-1}(\rho_x+\rho^2).   
\label{eq:temp} 
\eeq   
The density $\rho$, given by 
(\ref{eq:rho}), may be written as a formal 
series in powers of the spectral parameter 
$\lambda$, with coefficients determined 
recursively from (\ref{eq:spat}). Substituting a corresponding 
expansion for the current $j$ into (\ref{eq:temp}) and 
comparing powers of $\lambda$ yields an infinite sequence 
of conservation laws. However, it turns out that two 
different expansions are possible, leading to two infinite sequences  
of conserved quantities for (\ref{eq:tdnodisp}). 

With $m=-p^3$ as before, the first expansion takes the form 
\beq 
\rho=p\,\zeta^{-1}+\sum_{n=0}^{\infty}\rho^{(n)}\zeta^n, \qquad 
\lambda=\zeta^{-3}, \label{eq:exp1} 
\eeq 
with  the corresponding expansion of the current being 
\beq 
j=-pu\zeta^{-1}+\sum_{n=0}^{\infty}j^{(n)}\zeta^n. 
\eeq   
Thus the leading order ($\zeta^{-1}$ term) in (\ref{eq:temp}) is 
just the conservation law (\ref{eq:con3}). The first two  
densities and currents found recursively from 
(\ref{eq:spat}, \ref{eq:temp}) are 
\beq 
\begin{array}{ll} 
\rho^{(0)}=-p_x p^{-1}, &  
\rho^{(1)}=\frac{2}{3}p_{xx}p^{-2}-p_x^2p^{-3}+\frac{1}{3p}; \\ 
j^{(0)}=u_x-u\rho^{(0)}, & j^{(1)}=-p^2-u\rho^{(1)}. 
\end{array} 
\eeq 
For brevity we omit the recursion relations for 
$\rho^{(n)},j^{(n)}$, but note that the densities take the form 
$\rho^{(n)}=P^{(n)}/p^{2n+1}$ with $P^{(n)}$ a polyonmial 
of degree $n+1$ in $p,p_x,\ldots,p_{(n+1)x}$; the even terms 
$\rho^{(2n)}$ are exact $x$ derivatives. The densities $p$ 
and $\rho^{(1)}$ correspond to the conserved quantities 
$H_5$ and $H_7$ in (\ref{eq:hamils}) below. 

The second expansion is in positive powers of $\lambda$, viz 
\beq 
\rho=\sum_{n=0}^\infty r^{(n)}\lambda^{n+1}, 
\qquad j=\sum_{n=0}^\infty g^{(n)}\lambda^{n+1}. 
\eeq  
Again details of the recursive formulae for 
$r^{(n)},g^{(n)}$ are omitted,  but we note that the odd  
densities $r^{(2n+1)}$ are all exact derivatives, and 
apart from 
\beq 
r^{(0)}=u, \qquad r^{(2)}=u^3 
\eeq 
(which yield the conserved quantities $H_0$ and 
$H_{-1}$ below) all other non-trivial densities in this 
sequence are non-local since they arise by repeated application 
of the inverse of the Helmholtz operator 
$\mathcal{L}=(1-\partial_x^2)$   
appearing on the left hand side of (\ref{eq:spat}). 

There is another conservation law which does not appear in the 
above sequences, namely 
\beq 
\Big(v_{xx}^2+5v_x^2+4v^2\Big)_t 
=\left(4u^2v-\frac{2}{3}u^3-4v 
\mathcal{L}^{-1}(u^2)-v_x\mathcal{L}^{-1}(u^2)_x\right)_x, 
\label{eq:vcon} 
\eeq 
where $v$ is defined by  
\beq 
v:=(4-\partial_x^2)^{-1}u. 
\label{eq:vdef} 
\eeq    
This conservation law yields the conserved quantity 
$H_1$ in (\ref{eq:hamils}) below, since the density on the 
left hand side of (\ref{eq:vcon}) 
differs from the compact expression $mv$ by a total derivative.   
     
\section{Hamiltonian structures} 

\setcounter{equation}{0}

As derived above, the equation (\ref{eq:tdnodisp}) has 
an infinite sequence of conservation laws. Here we list some of the 
simplest associated conserved quantities: 
\beq 
\begin{array}{cclccl} 
H_{-1} & = & -\frac{1}{6}\int u^3\, dx, & 
H_{0} & = & -\frac{9}{2}\int m\, dx, \\  
H_1& =& \frac{1}{2}\int mv\, dx, &
H_5 & = & \int m^{1/3}\, dx, \\ 
H_7 & = & -\frac{1}{2}\int ( m_x^2 m^{-7/3} +9 m^{-1/3})\, dx. & &  &  
\end{array} 
\label{eq:hamils} 
\eeq   
In the above we take $u=(1-\partial_x^2)^{-1}m$ as before, 
and from (\ref{eq:vdef}) we have  
$v=(4-\partial_x^2)^{-1} 
(1-\partial_x^2)^{-1}m$. The labelling is such that 
$H_k$ generates a flow of weight $k$ with Hamiltonian 
operator $B_0$ 
as in (\ref{eq:hamops}) below.  

If we introduce the skew-symmetric differential operators 
$B_0, B_1$ according to 
\beq 
B_0=\partial_x(1-\partial_x^2)(4-\partial_x^2), 
\quad 
B_1=m^{2/3}\partial_xm^{1/3}(\partial_x-\partial_x^3)^{-1} 
m^{1/3}\partial_xm^{2/3}, 
\label{eq:hamops} 
\eeq 
then we can immediately write the equation (\ref{eq:tdnodisp}) 
in terms of the gradient of a conserved quantity in two 
different ways, namely: 
\beq 
m_t=B_0\, \frac{\delta H_{-1}}{\delta m} = 
B_1\, \frac{\delta H_{0}}{\delta m}. \label{eq:biham} 
\eeq 
The identity (\ref{eq:biham}) is our proposed bi-Hamiltonian 
form of the equation. However, in order to assert that 
the pair $B_0, B_1$ define a genuine bi-Hamiltonian 
structure it is necessary to check that both are 
Hamiltonian (Poisson) operators, and that they are 
compatible \cite{olver, dorf}. The first operator 
$B_0$ is constant coefficient and so the Jacobi identity is 
trivial and it is clearly Poisson, but for the non-local 
operator $B_1$ we have not succeeded in verifying the Jacobi 
identity. 

We have noted other properties of the operator pair 
(\ref{eq:hamops}) which strongly suggest that they do define 
the correct bi-Hamiltonian structure for (\ref{eq:tdnodisp}). 
In the list (\ref{eq:hamils}) we see that $H_0$ is a Casimir 
for $B_0$ and $H_5$ is a Casimir for $B_1$. 
The translational flow in the hierarchy is 
\beq 
m_x=B_0\, \frac{\delta H_{1}}{\delta m} =
B_1\, \frac{\delta H_{7}}{\delta m}, \label{eq:transl}
\eeq
where the second equality in (\ref{eq:transl}) 
requires suitable definition of the integral operator 
$\partial_x^{-1}$ in  
$B_1$. Also, the fifth-order vector field 
$B_0\, \frac{\delta H_{5}}{\delta m}$ is just the reciprocal 
transformation of the fifth-order Kaup-Kupershmidt equation. This 
is strong evidence that the operator $B_0 \, B_1^{-1}$ 
should be the reciprocal transform of the Kaup-Kupershmidt 
recursion operator (\ref{eq:kkrec}).

\section{Peakon dynamics} 

\setcounter{equation}{0}

Here we consider the dynamics of multi-peakon solutions 
of the equation (\ref{eq:tdnodisp}), 
which take the form 
\beq 
u(x,t)=\sum_{j=1}^Np_j(t)e^{-|x-q_j(t)|}, \quad 
m=2\sum_{j=1}^Np_j(t)\delta (x-q_j(t)),  
\label{eq:peaky} 
\eeq 
with N, the number of peakons, being arbitrary.  
 
Substituting the expression (\ref{eq:peaky}) into the equation 
(\ref{eq:tdnodisp}) yields the coupled ODEs for $q_j(t),p_j(t)$, 
namely   
\beq
\begin{array}{ccl}
\dot{p}_j &=&
2\,\sum_{k=1}^Np_j p_k\, sgn(q_j-q_k)e^{-|q_j-q_k|} ,
\\
\dot{q}_j &=&
\sum_{k=1}^Np_ke^{-|q_j-q_k|}, 
\end{array} 
\label{eq:tdODEeqn}
\eeq
which is just the system (\ref{eq:ODEeqn}) in the 
particular case $b=3$. Although the variables 
$q_j,p_j$ are not canonical, and we have not found a 
(non-canonical) Hamiltonian structure except in the 
special case $N=2$, the dynamical 
system (\ref{eq:tdODEeqn}) is integrable in the sense that 
it arises from a Lax equation 
\beq 
\dot{\mathsf{L}}=[\mathsf{M},\mathsf{L}] 
\eeq
where $\mathsf{L}$ and $\mathsf{M}$ take the form 
\beq 
\mathsf{L}=\frac{1}{2}(\mathsf{A}-\mathsf{A}^T-1-2\mathsf{C}) 
\mathsf{P}, \qquad 
\mathsf{M}=\mathsf{D}-(\mathsf{A}-\mathsf{A}^T)
\mathsf{P}. \label{eq:matrices} 
\eeq 

In order to specify the matrices in the Lax pair, which 
are built out of the strictly lower 
triangular matrices $\mathsf{A},\mathsf{C}$ and the diagonal matrices 
$\mathsf{P},\mathsf{D}$, we first note that an immediate consequence of 
the system (\ref{eq:tdODEeqn}) is that 
$\frac{d}{dt}sgn(q_j-q_k)=0$, 
so that the peakons preserve their relative ordering (as for 
Camassa-Holm peakons). Without loss of generality we take the ordering 
$q_1<q_2<\ldots<q_N$  
and then for the matrices appearing in (\ref{eq:matrices}), 
the non-zero components of $\mathsf{A}$ and 
$\mathsf{C}$ may be given (without modulus signs) as  
\beq 
A_{jk}=e^{q_k-q_j}, \quad 
C_{jk}=1, \quad j>k, 
\eeq 
while the diagonal elements of $\mathsf{P},\mathsf{D}$ are 
the components of the vectors 
\beq 
\bf{p}={\it (p_1,p_2,\ldots,p_N)^T}, \qquad 
\bf{d}=(\mathsf{A}-\mathsf{A}^{{\it T}})\bf{p} 
\eeq 
respectively. 

\begin{figure}[ht!]
\centerline{
\scalebox{0.45}{\includegraphics{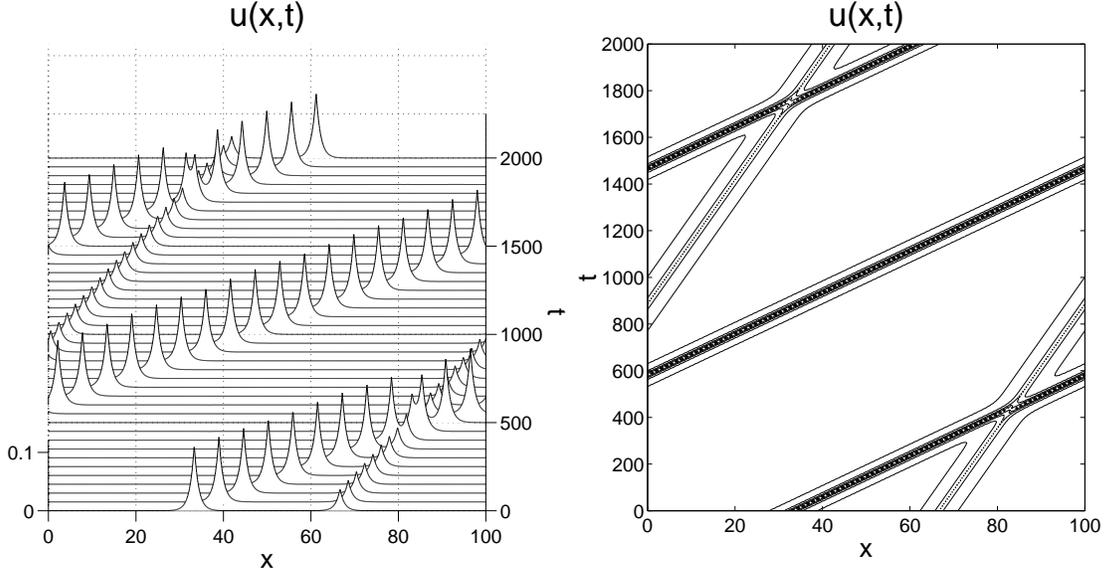}}
}
\caption{Two rear-end collisionss of $b=3$ peakons.   The initial positions
  are $x=33$ and $x=66$. 
  The faster peakon moves at three times the speed of the
  slower one.  For this ratio of speeds,
  both collisions result in a phase shift to the right for the faster
  space-time trajectory, but no phase shift for the slower one.}
\label{rear_collision-waterfall/contour_fig}
\end{figure}

In the case of 2-peakon scattering, i.e. the equations 
(\ref{eq:tdODEeqn}) for N=2, we consider 
two peakons that are initially
well separated with $q_1<q_2$ and have asymptotic speeds $c_1$ and $c_2$ 
as $t\to-\infty$ with $c_1>c_2$ and $c_1>0$ so that they eventually
collide. The situation is shown in Figure
\ref{rear_collision-waterfall/contour_fig}. 
 

In this case, we have two
conserved quantities 
\beq 
P:=p_1+p_2 =c_1+c_2, 
H:=p_1^2 + p_2^2
+
2p_1p_2\,
e^{q_1-q_2}\big(2-e^{q_1-q_2}\big)
=c_1^2+c_2^2.  
\eeq               
The equations of motion are then most easily solved in terms of the 
sum and difference variables 
$P$, $Q=q_1+q_2$,
$p=p_1-p_2$ and $q=q_1-q_2$, and the quadratures involved 
are almost the same as 
in the Camassa-Holm case (see \cite{ch2}). For the positions of the 
peaks we find the explicit formulae 
\begin{eqnarray}
q_1(t)
&=&
c_1\,t
+ \frac{1}{2}\log(4\Gamma(c_1-c_2)^2)
- \log\Big(\Gamma e^{(c_1-c_2)\,t}+4c_2P\Big)
\,,\label{q1-soln}\\
q_2(t)
&=&
c_2\,t
- \frac{1}{2}\log(4\Gamma(c_1-c_2)^2)
+ \log\Big(\Gamma e^{(c_1-c_2)\,t}+4c_1P\Big)
\,,\label{q2-soln}
\end{eqnarray}
where $\Gamma$ is an integration constant. 

In the limit $t\to\infty$, these formulas show that the solitons exchange
their asymptotic speeds, or equivalently  their momenta and amplitudes, as
\begin{equation}
\lim_{t\to\infty}q_1(t)=c_2t
\quad\hbox{and}\quad
\lim_{t\to\infty}q_2(t)=c_1t
\,.
\end{equation}
Thus, the main effect in the peakon scattering is an exchange of momentum,
or amplitude, between the two peakons, resulting only in a phase shift at
asymptotic times. The phase shift for the faster soliton (the one with
speed  $c_1$ in the limit $t\to-\infty$) is defined and evaluated using
(\ref{q1-soln}) and (\ref{q2-soln}) as
\begin{equation}\label{phaseshift-fast}
\Delta q_f = q_2(+\infty)-q_1(-\infty)
=\log\bigg[\frac{c_1(c_1+c_2)}{(c_1-c_2)^2}\bigg]
\,.
\end{equation}
Likewise, the phase shift for the slower soliton (the one with speed
$c_2$ in the limit $t\to-\infty$) is defined and evaluated using
(\ref{q1-soln}) and (\ref{q2-soln}) as
\begin{equation}\label{phaseshift-slow}
\Delta q_s = q_1(+\infty)-q_2(-\infty)
=
\log\bigg[\frac{(c_1-c_2)^2}{c_2(c_1+c_2)}\bigg]
\,.
\end{equation}
So for $c_1/c_2>3$ both peakons experience a forward shift, and 
for $1<c_1/c_2<3$ the faster peakon shifts forward while the slower one
shifts backward; 
the case $c_1/c_2=3$ is the turning point at which the slower peakon has
no phase shift. Remarkably, phase shift scattering rules corresponding 
precisely to these hold for Camassa-Holm peakons, except with $3\to2$ 
\cite{ch2}. 

An interesting special case is the peakon-antipeakon collision when 
$c_1$ and $c_2$ have opposite signs; only in this case can the 
solitons overlap when the variable $p=p_1-p_2$ diverges. For the 
perfectly antisymmetric collision $c_1=-c_2=c$ the resulting 
solution of the  partial differential equation (\ref{eq:peak}) is 
\beq 
\begin{array}{l} 
u(x,t)=
\frac{c}{1-e^{-2c|t|}}
\Big[e^{-|x+c|t|\,|}
-
e^{-|x-c|t|\,|}\Big], \\ 
m(x,t)= 
\frac{2c}{1-e^{-2c|t|}}
\Big[\delta(x+c|t|)
-
\delta(x-c|t|)\Big]. \end{array}  
\eeq 

\begin{figure}[ht!]
\centerline{
\scalebox{0.45}{\includegraphics{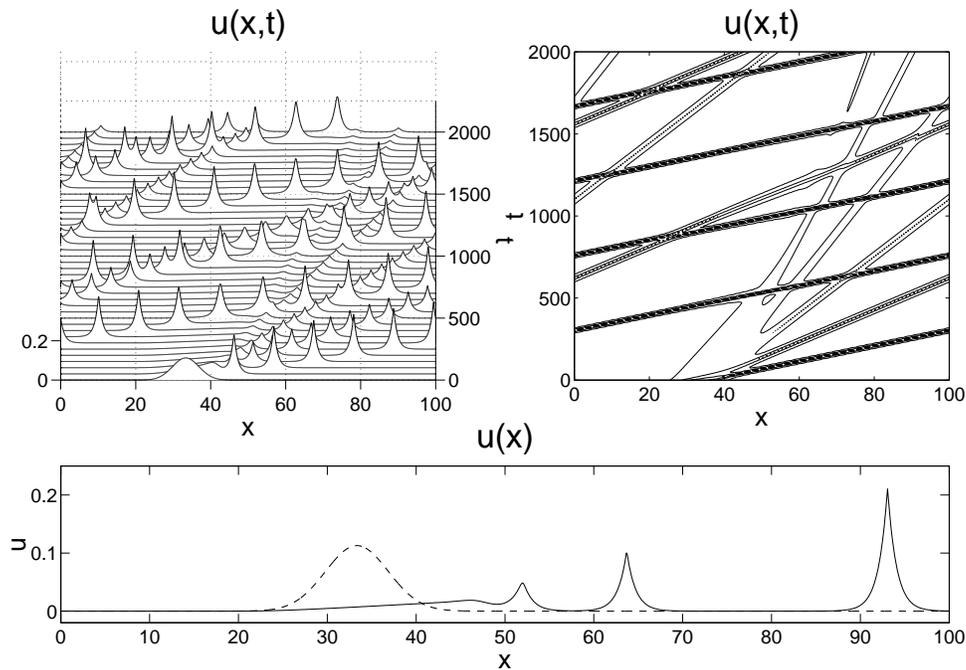}}
}
\caption{$b = 3$ peakons emerging from a Gaussian of unit
  area and $\sigma = 5$ centered about $x = 33$ on a periodic domain of
  length $L = 100$.  The fastest peakon recrosses the domain five times and
  has many elastic interactions with the slower ones.}
\label{Gauss_InitialCond-waterfall/contour_fig}
\end{figure}

In numerical simulations, we also investigated the emergence of peakons
from a Gaussian initial condition. The integrable behavior is
evidenced in Figure \ref{Gauss_InitialCond-waterfall/contour_fig} as the
peakons collide elastically as they recross the periodic domain.


\section{Conclusions}

In a forthcoming article \cite{dhh} we will consider an integro-differential 
generalization of (\ref{eq:bfamily}), extending the results of 
\cite{fringer} to include an extra parameter $b$, such that there 
are multi-pulson solutions described by the finite-dimensional 
system (\ref{eq:ODEeqn}) with the generating function        
\begin{eqnarray}
G_N = \frac{1}{2}\sum_{i,j=1}^{N} p_i p_j\, g(q_i-q_j)
\,.\nonumber
\end{eqnarray}
for an arbitrary even function $g(x)$ (so that $g(x)=e^{-|x|}$ in the 
peakons case). It turns out that for arbitrary $b$ values the 
two-pulson ($N=2$)
dynamics is Hamiltonian and integrable.  

\section{Acknowledgements}

We would like to thank the Isaac Newton Institute, Cambridge
for supporting our participation in the NEEDS workshop 
and the Programme on Integrable Systems. We are grateful to 
M. Staley for making the figures for us using his PDEUM code.


\small

\begin{thebibliography}{99}



\bibitem{ch}R.~Camassa and D.D.~Holm,
Phys. Rev. Lett. 71 (1993) 1661-1664.

\bibitem{ch2}R.~Camassa, D.D.~Holm and J.M.~Hyman, 
Advances in Applied Mechanics 31 (1994) 1-33.   

\bibitem{dhh}A.~Degasperis, A.N.W.~Hone and D.D.~Holm, 
{\it A Class of Equations with Peakon and
Pulson Solutions}, in preparation. 

\bibitem{dega}A. Degasperis and M. Procesi, Asymptotic integrability, in
{\it Symmetry and Perturbation Theory}, edited by A. Degasperis and G.
Gaeta, World Scientific (1999) pp.23-37.


\bibitem{bull}R.K.~Dodd and R.K.~Bullough,
Proc. Roy. Soc. Lond. A 352 (1977) 481.

\bibitem{dorf}I.~Dorfman, 
{\it Dirac Structures and Integrability of Nonlinear Evolution Equations}, 
John Wiley \& Sons (1993). 

\bibitem{DGH[2001]}
H. Dullin, G. Gottwald and D. D. Holm,
Phys. Rev. Lett. 87 (2001) 194501-1.

\bibitem{fringer}O.~Fringer and D.D.~Holm,
Physica D 150 (2001) 237-263.

\bibitem{fuchss}B.~Fuchssteiner, 
Physica D 95 (1996) 229-243.  

\bibitem{pick}C.~Gilson and A.~Pickering,
J. Phys. A 28 (1995) 2871-2888.

\bibitem{pick2}P.R.~Gordoa and A.~Pickering,
J. Math. Phys. 40 (1999) 5749-5786.

\bibitem{hone1}A.N.W.~Hone, 
J. Phys. A 32 (1999) L307-314.

\bibitem{hone2}A.N.W.~Hone,
Phys. Lett. A 263 (1999) 347-354.

\bibitem{hone3}A.N.W.~Hone,
Applied Mathematics Letters 13 (2000) 37-42. 

\bibitem{rogers1}J.G.~Kingston and C.~Rogers,
Phys. Lett. A 92 (1982) 261-264.


\bibitem{kraenkel}R.A.~Kraenkel and A.~Zenchuk, 
Phys. Lett. A 260 (1999) 218-224. 

\bibitem{olver}P.~Olver and P.~Rosenau, 
Phys. Rev. E 53 (1996) 1900-1906. 



\bibitem{weak}A.~Ramani, B.~Dorizzi and B.~Grammaticos,
Phys. Rev. Lett. 49 (1982) 1538-1541.


\bibitem{rogers3}C.~Rogers, {\it Reciprocal Transformations
and Their Applications},
in {\it Nonlinear Evolutions}, 109-123,
ed. J.~Leon, World Scientific, Singapore (1988).

\bibitem{rogers2}C.~Rogers,
{\it B\"{a}cklund transformations in soliton theory},
in {\it Soliton theory: a survey of results},
97-130, ed. A.P.~Fordy, Manchester
University Press (1990).

\bibitem{tzit}G.~Tzitzeica, 
C.R. Acad. Sci. T.150 (1910), 955-956; 
C.R. Acad. Sci. T.150 (1910), 1227-1229. 

\bibitem{wtc}J.~Weiss, M.~Tabor and G.J.~Carnevale,
J. Math. Phys. 24 (1983) 522-526.


\end{thebibliography}
\end{document}